\documentstyle[aps,preprint]{revtex}
\newcommand{\be}{\begin{equation}}
\newcommand{\ee}{\end{equation}}
\newcommand{\bea}{\begin{eqnarray}}
\newcommand{\eea}{\end{eqnarray}}

\def\6{\partial}
\def\tp{{\tilde{\phi}}^{(\mu )}}
\def\F{{\cal{F}}^{(\mu )}}
\def\p{{\phi}^{(\mu )}}
\def\G{{\cal{G}}^{(\mu )}}
\draft
\begin{document}

\title{Duality of Coordinates and Matter Fields in Curved Spacetime}
\author{M. A. De Andrade \footnote{e-mail: marco@cbpf.br} }
\address{ Grupo de F\'{\i}sica Te\'{o}rica, 
Universidade Catolica de Petr\'{o}polis (UCP)\\
Rua Bar\~{a}o do Amazonas 124, 25.685-000 Petr\'{o}polis-RJ, Brazil}
\author{I. V. Vancea\footnote{e-mail: ivancea@ift.unesp.br} }
\address{ Instituto de F\'{\i}sica Te\'{o}rica, Universidade 
Estadual Paulista (UNESP)\\
Rua Pamplona 145, 01405-900 S\~{a}o Paulo-SP\\
and\\
Departamento de F\'{\i}sica Te\'{o}rica,
Universidade do Estado do Rio de Janeiro (UERJ)\\
Rua S\~{a}o Francisco Xavier 524, Maracan\~{a}, Rio de Janeiro-RJ, Brazil}

\maketitle

\pacs{02.40.-k, 11.10.+t, 03.70.+k}

\begin{abstract}
We show that there exists a duality between the local coordinates and the
solutions of the Klein-Gordon equation in curved spacetime in the same sense
as in the Minkowski spacetime. However, the duality in curved spacetime does
not have the same generality as in flat spacetime and it holds only if the
system satisfies certain constraints. We derive these constraints and the
basic equations of duality and discuss the implications in the quantum
theory.

\end{abstract}


\vspace{1cm}

According to \cite{fm}, there exists a duality between the Cartesian
coordinates of flat spacetime and the solutions of the equation of
motion of a certain physical systems. In the nonrelativistic case, this
duality allows to express the coordinates as functionals on the wave
functions of an one particle quantum system and some other functionals
called prepotentials introduced in \cite{sw}(see also \cite{bm,mm}.) In
Minkowski spacetime {\em all} the Cartesian coordinates are functionals on
the solutions of the classical Klein-Gordon equation and prepotentials.

The results from \cite{fm} suggest that the spacetime measurements can
be represented in terms of either coordinates or scalar fields, maybe even
quantum fields. This interpretation might be particularly useful
at the Planck scale where the classical rulers and clocks can be at most
idealized while the fields are physical. Also, the Heisenberg inequalities
introduce a strong indeterminancy in the spacetime measurements performed
by the inertial observers \cite{dfr,jm}. In particular, this might be a
hint that the microscopic structure of spacetime is actually operatorial
and noncommutative \cite{cc,cl}. However, at distances of the order of
Planck length, the effects of gravity are rather strong. According to
the principles of general relativity, in the presence of gravity the
coordinates cease to have the same important role as in the
flat spacetime physics\cite{rw}. Nevertheless, they are still essential to
perform local physical measurements. Therefore, it is important to
understand whether the coordinate-field duality described in \cite{fm} has
a counterpart in curved spacetime.

The purpose of this letter is to investigate the circumstances which allow
a duality between the local coordinates and the solutions of the Klein-Gordon
equation in curved spacetime in both classical and quantum theories (for
recent attempts to formulate the gravity in terms of quantum quantities see
\cite{lr}-\cite{pv}.) The sought for duality relations should have the same
interpretation as in the flat spacetime and should include it as a particular
case.

Let us consider a curved spacetime manifold $M$ of dimension $n$, endowed with
a Lorentzian metric tensor field $g$ and a scalar field $\phi$ of mass $m$. In
a local coordinate system $(x^0,x^1,\ldots ,x^{n-1})$ the equation of motion
of $\phi$ is given by
\be
(\Box_x +m^2 +\xi R(x))\phi (x)=0,
\label{kgeq}
\ee
where $\Box_x$ is the d'Alambertian, $\xi$ is a real parameter and $R(x)$
is the scalar curvature of $M$ at $x$. 
In order to find a duality of the same type as in \cite{fm} one has to express
the coordinate along a direction $x^{\mu}$ as a function of some 
solution of (\ref{kgeq}) for which $x^{\nu}$ (with $\nu \neq \mu$) represent
independent parameters. These "one dimensional" solutions do not exist in
general unless some constraints are imposed on the system. In order to
find them we put (\ref{kgeq}) under the form
\be
M^{(\mu \mu )}\6_{\mu}\6_{\mu}\phi(x^{\mu}) +
N^{(\mu )}\6_{\mu}\phi(x^{\mu}) +
P^{(\mu)}\phi(x^{\mu}) + Q^{(\mu )}=0
\label{baseq}
\ee
for $\mu = 0, 1, \ldots , n-1$. The coefficients $M^{(\mu \mu )}$,
$N^{(\mu )}$, $P^{(\mu)}$ and $Q^{(\mu )}$ are specific functions on $g$, $\xi$
and $R$, but it is convenients to treat (\ref{baseq}) in its full generality
and to imposed the dependence later as a constraint. Since the variable in
(\ref{baseq}) is $x^{\mu}$ while $x^{\nu}$ are treated as parameters for
$\nu \neq \mu$ and since we are seeking for independent dualities along
each $\mu$, we have to make a first technical assumption that $\phi$ and
$\6_{\nu}\phi$ can be treated as independent functions. As a consequence,
the coefficients in (\ref{baseq}) can be considered as continuous functions
on $x^{\mu}$ depending on the parameteres $x^{\nu}$. The second nontrivial 
requirement that should be satisfied by the system is that (\ref{baseq})
admits two linearly independent solutions.  This condition can be translated
into a statement about a second order differential equation obtained by
changing the coordinates in (\ref{baseq}) to local Cartesian coordinates in
$R^{n}$ \cite{mav}. A general solution of the later can be constructed out
of a particular integral of (\ref{baseq}) and a complementary function that
is a solution of the corresponding homogeneous equation. For the later a
fundamental solution always exist \cite{math}.

The two conditions above refer to the possibility of decomposing 
(\ref{baseq}) along the directions $x^{\mu}$ as in the flat case. In general,
it is known that such of decomposition is not possible due to the presence of 
the off-diagonal elements of the metric in the d'Alambertian. Moreover, even 
the notion of direction is of little use since there is no invariance of the 
metric along any vector field \cite{rw}. These facts question the existence of
any solution to the conditions stated above. However, a large class of manifolds 
should meet these requirments, namely the manifolds that present at least a
number of local Killing vectors equal to the dimension of the manifold. In this 
case it is known that the Klein-Gordon equation admits wave-function solutions
along the integral lines of the Killing vectors \cite{rw}. Then the fundamental 
theorem for linear differential equations \cite{math} guarantees that the
homogeneous one-variable equation (\ref{baseq}) has two linearly independent 
solutions and that the inhomogenous equation has solution over the interval
where the coeficients $M^{(\mu \mu )}$, $N^{(\mu )}$, $P^{(\mu)}$ and 
$Q^{(\mu )}$ are continuous and defined and $M^{(\mu \mu )} \neq 0$. The 
existence of the second linearly independent solution depends on the specific 
form
of $Q^{(\mu )}$ which at its turn depends on the particular metric. The fact
that $\phi$ and $\6_{\nu}\phi$ are treated as independent functions implies
that any duality relation that can be derived under this assumption is strictly
local. Otherwise, a relationship between the two functions can be shown to 
arise.

Let us assume that the above requirements are satisfied for our system. We 
denote by $\phi^{\mu}$ and
${\tilde{\phi}}^{\mu}$ two linearly independent solutions of (\ref{baseq}).
Following \cite{fm} we introduce the prepotentials $\F [\phi^{\mu}]$ by
\be
\tp = \frac{\6 \F [\phi^{\mu}]}{\6 \p}.
\label{prep}
\ee
The variation of $\F$ with respect to $x^{\mu}$ is given by
\be
\6_{\mu}\F = \frac{1}{2}\6_{\mu}(\tp \p ) + \frac{1}{2} W^{(\mu )},
\label{derprep}
\ee
where
\be
W^{(\mu )}= \tp \6_{\mu} \p - \6_{\mu} \tp \p
\label{wronsk}
\ee
is the Wronskian of $\p$ and $\tp$. In general, $W^{(\mu )}$ is a continuous
functions on $x^{\mu}$ and depends on the parameters $x^{\nu}$ with
$\nu \neq \mu$. However, since we want to express $x^{\mu}$ as an explicit
function on $\p$ and $\tp$ from (\ref{derprep}) $W^{(\mu )}$ should be a
(nonvanishing) constant function with respect to $x^{\mu}$. This imposes the
following supplementary constraint on the system
\be
N^{(\mu )}W^{(\mu )} + M^{(\mu \mu )}Q^{(\mu )}(\tp -\p )=0,
\label{constr1}
\ee
for $\mu = 0, 1, \ldots , n-1$. Now integrating (\ref{derprep}) with
respect to $x^{\mu}$ we obtain the following relation
\be
x^{\mu} = \frac{2}{W^{(\mu )}}(\F - \frac{1}{2}\tp\p -C^{(\mu )})
\label{duality}
\ee
for $\mu = 0, 1, \ldots ,n-1$. Here, $C^{(\mu )}$ is an integration constant
with respect to $x^{\mu}$ and an arbitrary function on the parameters.
Equation (\ref{duality}) expresses the duality between the coordinate
$x^{\mu}$ on one hand, and $\p$, $\tp$ and $\F$ on the other hand. We note
that it has the same form as in flat spacetime, due to the imposed
requirements. By setting $W^{(\mu )}$ and $C^{(\mu )}$ constants with
respect to $x^{\nu}$ as in the flat spacetime and by using (\ref{prep}) one
can see that $x^{\mu}$ depends explicitely only on $\p$ and $\F$ which is
unknown.

In order to derive the equation satisfied by $\F$ we introduce the
functional $x^{\mu} \equiv \G [\p ]$. Follwing \cite{mm} we express the
derivatives with respect to $x^{\mu}$ in terms of the derivatives with respect
to $\p$. We see that
\be
\frac{\6}{\6 x^{\mu}} = (\6 \G )^{-1}\frac{\6}{\6 \p}
\label{firstder}
\ee
and
\be
\frac{\6^2}{\6 x^{\mu} \6 x^{\mu}} =
(\6 \G )^{-1} [ -(\6 \G )^{-2}(\6^2 \G )\frac{\6}{\6 \p} +
(\6 \G )^{-1}\frac{\6^2}{\6 \p \6 \p}],
\label{secder}
\ee
where $\6 = \6 / \6 \p $. Using (\ref{duality}) we calculate the first
two derivatives of $\G$ and find the following relations
\be
\6 \G = \frac{1}{W^{(\mu )}}(\6 \F - \6^2 \F \p ) ~~, ~~
\6^2 \G = -\frac{1}{W^{(\mu )}}\6^3 \F \p .
\label{derG}
\ee
Next, by using (\ref{prep}), (\ref{firstder}), (\ref{secder}) and
(\ref{derG}) in (\ref{baseq}) we obtain after some simple computations the
following equation
\bea
M^{(\mu \mu )} \6^3 \F \6 \F & + &
\frac{1}{W^{(\mu )}} N^{(\mu )}(\6 \F - \6^2 \F \p )^2 \6^2 \F \nonumber\\
&+& \frac{1}{W^{(\mu )}}(P^{(\mu )} \6 \F + Q^{(\mu )})(\6 \F -
\6^2 \F \p )^3 =0
\label{eqforF}
\eea
for $\mu = 0, 1, \ldots , n-1 $. Equation (\ref{eqforF}) is the
"equation of motion" of $\F$ in the space of solutions of (\ref{baseq}).
Together with (\ref{duality}) it represents the generalization of the
duality relations obtained in \cite{fm} to a curved manifold. However,
(\ref{duality}) and (\ref{eqforF}) do not have the same generality as in the
Minkowski space since we have assumed that the system satisfies three strong
mathematical requirements.
Actually, in order to make contact with physics, the general coefficients
of (\ref{baseq}) are connected to the geometrical objects defined on $M$
through the following relations
\bea
M^{(\mu \mu )} & = & g^{\mu \mu } \sqrt{g}~~
,~~N^{(\mu )}=\6_{\mu}(g^{\mu \mu }\sqrt{g}) +
\sum_{\nu \neq \mu}\6_{\nu }(g^{\nu \mu} \sqrt{g})\nonumber\\
P^{(\mu )} & = & \sqrt{g}(m^2 + \xi R) ~~,~~
Q^{(\mu )}=2\sum_{\nu \neq \mu}(g^{\mu \nu } \sqrt{g} \6_{\mu} f_{\nu} +
\frac{1}{2}\6_{\mu}(g^{\mu \nu }\sqrt{g})f_{\nu })\nonumber\\
f_{\nu} & = & \6_{\nu}\phi
\label{coeff}
\eea
for $\mu = 0,1,\ldots,n-1$. Thus we see that (\ref{constr1}) represents a
constraint on the geometry as well as on the derivatives of $\phi$ with
respect to the parameters $x^{\nu}$ with $\nu \neq \mu$.

Our technical assumptions on the non-homogeneous second order differential
equation (\ref{baseq}) are quite restrictive. Nevertheless, they include
the homogeneous equation $Q^{(\mu )} = 0$. The relations describing the
duality can be directly obtained from the general case. Indeed, from
(\ref{coeff}) we see that the homogeneous case is obtained for the metrics
on $M$ that satisfy the following relation
\be
\sum_{\nu \neq \mu } (g^{\mu \nu}\sqrt{g}\6_{\mu}\6_{\nu} + \frac{1}{2}
(\6_{\mu}(g^{\mu \nu}\sqrt{g})\6_{\nu})\phi = 0,
\label{constr2}
\ee
which represent just the homogenity condition. Also, the condition 
(\ref{constr1}) of having a constant Wronskian becomes in this case 
\be
\sum_{\nu = 0}^{n-1}\6_{\nu}(g^{\nu \mu}\sqrt{g})=0,
\label{constr3}
\ee
which can be obtained from (\ref{constr1}) and (\ref{coeff}) with the 
homogenity condition (\ref{constr2}). The equation (\ref{constr3}) 
fixes the metric on $M$. When used together with (\ref{constr2}) in the
Klein-Gordon equation it leads to the following equation for 
$\phi$ along $x^{\mu}$
\be
\frac{1}{2}\6_{\mu}(g^{\mu \mu}\sqrt{g})\6_{\mu}\phi +
g^{\mu \mu}\sqrt{g}\6_{\mu}\6_{\mu}\phi + (m^2 + \xi R(x))\phi = 0 .
\label{eqplus}
\ee
Equation (\ref{eqplus}) is a homogeneous differential equation of rank
two and it admits always two linearly independent solutions if the
coefficients are continuous and nonsingular. The duality relation
(\ref{duality}) has the same form in the homogeneous case, but the equation
satisfied by $\F$ takes the following form
\be
M^{(\mu \mu )}\6^3 \F + \frac{1}{W^{(\mu )}}P^{(\mu )}(\6 \F - \6^3 \F \p )^3 =0.
\label{homF}
\ee

In particular, the Minkowski case studied in \cite{fm} can be obtained from
the general nonhomogeneous theory by setting
\be
V^{(\mu )}= \frac{1}{\tp} Q^{(\mu )}
\label{mink}
\ee
and by choosing appropriate normalization conditions for $\p$. The
corresponding duality relations can be easily deduced from
(\ref{duality}) and (\ref{eqforF}).

Let us discuss briefly the coordinate-field duality in the quantum theory
(for more details see \cite{mav}). In the operatorial formalism, the
general solution of (\ref{kgeq}) can be expanded as
\be
\phi (x) = \sum_{\alpha }(a_{\alpha }f_{\alpha}(x) + a_{\alpha}^{\dagger}
\bar{f_{\alpha}}(x)),
\label{expand}
\ee
where $\{f_{\alpha}(x)\}$ is a local orthonormal and complete set of
solutions of
(\ref{kgeq}) and $a_{\alpha}$ and $a_{\alpha}^{\dagger}$ are the
(covariant) annihilation and creation operators, respectively. Since the
Poincar\'{e} group is not a symmetry group of the system, the particle
interpretation is problematic and the theory suffers from the known
inconsistencies \cite{rw}. Nevertheless, let us examine the circumstances
under which this theory can display a coordinate-field duality of the
same kind as the one discussed in \cite{fm}. As in the classical case,
in general there is no such of duality in the system. However, 
one can see that it exists if the Klein-Gordon field admits
a local decomposition in independent modes along each of the directions
$\mu = 0, 1, \ldots , n-1$. In particular, this implies that the
equation (\ref{baseq}) has an orthonormal and complete set of solutions
$\{ f_{\alpha}^{(\mu )} \}$, where $\alpha$ represent all the necessary
indices to label an independent mode. This linearizing condition strongly 
constraints the
possible types of coefficients to be used in the quantum equation
(\ref{baseq}) and consequently it constraints the spacetime geometries.

Let us assume that a $\{ f_{\alpha}^{(\mu )} \}$ exists such that it satisfies the
following conditions
\bea
f_{\alpha}^{(\mu )} \cdot f_{\beta}^{(\nu )} &=&
- {\bar{f}}_{\alpha}^{(\mu )} \cdot {\bar{f}}_{\beta}^{(\nu )} =
\delta^{\mu \nu }\delta_{\alpha \beta}
\nonumber\\
f_{\alpha}^{(\mu )} \cdot {\bar{f}}_{\alpha}^{(\mu )} &=&0
\label{scalprod}
\eea
where "$\cdot$" is the scalar product in the space of solutions defined by an
integral over a spacelike hypersuface, then we can construct two linearly
independent solutions of (\ref{baseq}) of the following form
\be
\p_{\alpha}(x) = a_{\alpha} f_{\alpha}^{(\mu )}
~~,~~
\tp_{\alpha} (x) = a_{\alpha}^{\dagger} {\bar{f}}_{\alpha}^{(\mu )} ,
\label{quantsol}
\ee
which correspond to operators that locally annihilate and create,
respectively, the quantum mode $\alpha$ of the field along the direction
$\mu$. Using the same definitions (\ref{prep}) as in the classical case,
one can associate to each mode $\alpha$ the quantum prepotential
$\F_{\alpha}$ which is an operator depending on $\p_{\alpha}$.

The existence of a basis (\ref{scalprod}) restricts the possible
manifolds that can support a quantum duality of the type presented in 
\cite{fm}. In particular, let us suppose that $M$ and its geometric 
structure satisfies all the requirements necessary for a classical duality
between coordinates and fields to hold. As we argued previously, depending on
$Q^{(\mu )}$, this can take place if there are local Killing vector fields that
define the local directions arround $(x^{\mu})$ on $M$. In this case, the
modes $f_{\alpha}^{(\mu )}$ can be taken as the eigenvalues of the translation
operator along the Killing vector that defines the direction $x^{\mu}$. In
particular this means that $\6_{\mu} f_{\alpha}^{(\mu )} = 
k_{\mu} f_{\alpha}^{(\mu )}$ and that the equation (\ref{baseq}) is also 
satisfied by the elements of the basis (for certain functions $Q^{(\mu )}$
the solutions of the inhomogeneous differential equation
(\ref{baseq}) can be decomposed in the modes of the translation 
operator, the trivial case being of the homogeneous equations.)
Now if we introduce a solution of the form
(\ref{quantsol}) in (\ref{baseq}) and expand the function $Q^{(\mu )}$ in the basis
(\ref{scalprod}) with the coefficients $q_{\alpha}^{\mu}$ 
then we obtain the following equation
\be
q_{\alpha}^{\mu} = -a_{\alpha}(k_{\mu}k_{\mu}M^{(\mu \mu )} + 
k_{\mu}N^{(\mu )} + P^{(\mu )}) \label{anon}.
\ee
Thus, a basis that satisfies (\ref{scalprod}) exists on $M$ if the
local directions are defined in terms of local Killing vectors and if
the classical duality coordinate-fields holds. Also, it is necessary that 
the coefficients of
(\ref{baseq}) that depend on the geometry of $M$ satisfy (\ref{anon}) beside
(\ref{constr1}). The resolution of the basis depends on $Q^{(\mu )}$. The 
equation (\ref{anon}) admits in general solutions, but these cannot be
uniquely determined only from it. However, beside this equation, the 
functions $M^{(\mu \mu )}$, $N^{(\mu )}$, $P^{(\mu)}$ and $Q^{(\mu )}$ should
also satisfy (\ref{constr1}) but even in this case the solutions are not 
completely determined even if their existence can be inferred from a simple
analysis of the rank of the system.
It is possible that a basis of the form (\ref{scalprod}) exists in a more
general or different case, but this problem is not clear to us at present.
The basis could be in principle continuous or discrete (for example in the
case of a compact Killing vector, the momentum eigenvalues are quantized.)

It is easy to verify that the rest of conditions that should be imposed on
the system actually refer only to the modes $f^{(\mu )}_{\alpha}$ and
they reduce to the ones discussed in the case of classical fields. This
is a consequence of
the fact that the modes are assumed to be independent. In particualar, the
coefficients $M^{(\mu \mu )}_{\alpha}$, $N^{(\mu )}_{\alpha}$ and
$Q^{(\mu )}_{\alpha}$ that enter the corresponding equation of the form
(\ref{baseq}) written for the mode $\alpha$ along the direction $\mu$ should
satisfy the constraint (\ref{constr1}) with $W^{(\mu )}_{\alpha}$ the
Wronskian of $f^{(\mu )}_{\alpha}$ and ${\bar{f}}^{(\mu )}_{\alpha} $.
Proceeding along the same line as in the
classical case, it turns out that the duality between coordinates and fields
is expressed by the following relation
\bea
X^{(\mu )}_{\alpha} & = & \frac{2}{W^{(\mu )}_{\alpha}}(\F_{\alpha} -
\frac{1}{2}{\bar{f}}^{(\mu)}_{\alpha} f^{(\mu )}_{\alpha} a_{\alpha}^{\dagger}
a_{\alpha} -
C^{(\mu)}_{\alpha} a_{\alpha}^{\dagger} a_{\alpha}),
\nonumber\\
X^{(\mu )}_{\alpha} & = & x^{\mu} a_{\alpha}^{\dagger} a_{\alpha}
\label{quantdual}
\eea
for $\mu = 0, 1, \ldots ,n-1$. Here, $W^{(\mu )}_{\alpha}$ and
$C^{(\mu)}_{\alpha}$ are constant functions with respect to $x^{\mu}$ and may
depend arbitrarily on the parameters $x^{\nu}$ with $\nu \neq \mu$. The
operators
$\F_{\alpha}$ are hemitean only if both of $W^{(\mu )}_{\alpha}$ and
$C^{(\mu )}_{\alpha}$ are real. There is an orthonormal and complete set of
states which are eigenstates of the operators $\F_{\alpha}$, $N_{\alpha}$
and $X^{(\mu )}_{\alpha}$ since
\be
[X^{(\mu )}_{\alpha} , X^{(\nu )}_{\beta}]=
[ \F_{\alpha}, {\cal{F}}^{(\nu )}_{\beta}]=
[X^{(\mu )}_{\alpha} , N_{\alpha}]=
[\F_{\alpha} , X^{(\nu )}_{\beta}]=
[\F_{\alpha} , N_{\alpha}]=0
\label{algop}
\ee
as a consequence of the independence of the modes.

If the operator $\6 X^{\mu }_{\alpha}$ is invertible it is possible to
obtain the equation of motion for $\F_{\alpha}$ in the space of modes.
Like in the classical case we first determine $\6 X^{\mu }_{\alpha}$
and $\6^2 X^{\mu }_{\alpha}$ and we obtain the following relations
\be
\6 X^{\mu }_{\alpha} = -\frac{2}{W^{(\mu )}_{\alpha}}[(\frac{1}{2} +
\frac{C^{(\mu )}_{\alpha}}
{ {\bar{f}}^{(\mu)}_{\alpha} f^{(\mu )}_{\alpha} })
\6^2 \F_{\alpha}\p_{\alpha} + (\frac{1}{2} -
\frac{ C^{(\mu )}_{\alpha} }{ {\bar{f}}^{(\mu)}_{\alpha} f^{(\mu )}_{\alpha}})
\6^2 \F_{\alpha}\p_{\alpha}]
\label{firstx}
\ee
and
\be
\6^2 X^{\mu }_{\alpha} = - \frac{2}{ W^{(\mu )}_{\alpha} }(\frac{1}{2} +
\frac{ C^{(\mu )}_{\alpha} }{ {\bar{f}}^{(\mu)}_{\alpha} f^{(\mu )}_{\alpha}})
\6^3 \F_{\alpha}\p_{\alpha} -\frac{4}{W^{(\mu )}}
(\frac{C^{(\mu )}_{\alpha}}{{\bar{f}}^{(\mu)}_{\alpha} f^{(\mu )}_{\alpha}})
\6^2 \F_{\alpha}.
\label{secondx}
\ee
Next introduce (\ref{firstx}) and (\ref{secondx}) in the operatorial equation
corresponding to (\ref{baseq}) and express the derivatives $\6_{\mu}$ in
terms of $\6 / \6 \p_{\alpha}$. For the sake of clarity let us introduce
the following notations for the functions that enter the final result
\bea
\Sigma &=& ({\bar{f}}^{(\mu)}_{\alpha} f^{(\mu )}_{\alpha} )^{-1}~~,
~~\Omega_1 = -\frac{2}{ W^{(\mu )}_{\alpha} }(\frac{1}{2} +
\frac{ C^{(\mu )}_{\alpha} }{ {\bar{f}}^{(\mu)}_{\alpha} f^{(\mu )}_{\alpha}})
\nonumber\\
\Omega_2 &=&- \frac{4}{W^{(\mu )}}
(\frac{C^{(\mu )}_{\alpha}}{{\bar{f}}^{(\mu)}_{\alpha} f^{(\mu )}_{\alpha}}).
\eea
Next we drop the indices $\mu$ and $\alpha$ and put a hat on operators. With
these notations the
equation satisfied by the quantum prepotentials has the following form
\bea
M \Sigma \6 \hat{\cal{F}}\hat{\phi}\hat{Y}[((\Sigma(\6^2\hat{\cal{F}}
\hat{\phi} &+&
\6\hat{\cal{F}})-\Sigma\6\hat{\cal{F}}\hat{\phi}\hat{Y}(\Omega_1 \6^3
\hat{\cal{F}}\hat{\phi} + \Omega_3 \6^2 \hat{\cal{F}}))\hat{Y} + 1)\6^2
 \hat{\cal{F}}
\nonumber\\
&+& \Sigma \6 \hat{\cal{F}}\hat{\phi}\hat{Y}\6^3 \hat{\cal{F}}] +
P \6\hat{\cal{F}} +Q=0,
\label{eqforFO}
\eea
where $\hat{Y}=(\6 \hat{X})^{-1}$. Equation (\ref{eqforFO}) is nonlinear
and nontrivial, however some simplifications are possible in the homogeneous
case and by taking the constant $C^{(\mu )}_{\alpha}=0$. Nevertheless,
even in the simplest case it is a very difficult task to find a nontrivial
solution of it.

The equation (\ref{quantdual}) and (\ref{eqforFO}) represent the content of
the duality between the coordinates and fields in the quantum theory. This
duality was obtained by imposing strong constraints on the system. The reason
for these constraints is the requirement that the duality has the same
interpretation as in the classical theory in the Minkowski spacetime
\cite{fm}(see also \cite{mav}.)

In conclusion, we have generalized the classical duality between coordinates
and matter fields to curved spacetime. In the classical case, in order to
obtain
a theory that reduces to the one discussed in \cite{fm},
the constraints expressed by (\ref{constr1}) and (\ref{coeff}) should be
imposed on the system. Moreover, in order to apply the method used to derive
the
flat spacetime duality, one should assume that the system satisfies some
restrictive mathematical conditions. In the quantum case, the duality exists
on those manifolds on which the local decomposition of the scalar fields in
independent modes is possible along each coordinate. We obtained in particular
the relations derived in the case of Minkowski spacetime in \cite{fm}. Our
analysis shows that the duality coordinate-fields should also hold locally 
in the case 
of other flat manifolds. However, we emphase that this is not a property
of general spacetime manifolds, at least not in the form of \cite{fm}. We
have shown that, in principle, this type of duality should hold for some
nontrivial manifolds but the complexity of the matematical restrictions that
should be imposed to the system prevents us of giving examples at present.

Despite the technical difficulties, the theory presented here might be useful
in studying the general relativity at Planck scale. One interesting way to
apply the coordinate-field duality to gravity would be to parametrize locally
the spacetime manifold in terms of matter fields in a consistent way with
the principles of general relativity.

\acknowledgements

We thank to S. Sorella and J. A. Helayel-Neto for important discussions.
M. A. A. is grateful to CBPF for support. I. V. V. thanks to CBPF and UCP 
for hospitality. I. V. V.'s work was partially supported by a FAPERJ and a 
FAPESP postdoc fellowships.

\end{document}